\documentclass[12pt,journal,onecolumn]{IEEEtran}
\usepackage{amsmath,amssymb,threeparttable,placeins,makecell,stfloats,cite}
\usepackage[dvips]{graphicx,epstopdf}
\usepackage[usenames]{color}
\usepackage{amsfonts}
\usepackage{latexsym}
\usepackage{setspace}
\usepackage{caption}
\usepackage{subcaption}
\usepackage{floatrow}
\newtheorem{theorem}{{{\textit{Theorem}}}}

\newtheorem{lemma}{{{\textit{Lemma}}}}

\newtheorem{property}{{{\textit{Property}}}}
\newtheorem{definition}{{{\textit{Definition}}}}

\newtheorem{remark}{{{\textit{Remark}}}}

\newtheorem{example}{{{\textit{Example}}}}

\newtheorem{construction}{{{\textit{Construction}}}}

\evensidemargin=\oddsidemargin
\begin{document}
\doublespacing{}
\title{New Constructions of Golay Complementary Pair/Array with Large Zero Correlation Zone}
\author{Zhi Gu,~\IEEEmembership{Student Member,~IEEE,}
Zhengchun Zhou,~\IEEEmembership{Member,~IEEE,}
    Avik Ranjan Adhikary,~\IEEEmembership{Member,~IEEE}
    Yanghe Feng,
    Pingzhi Fan,~\IEEEmembership{Fellow,~IEEE}
\thanks{Z. Gu is with the School of Information Science and Technology, Southwest Jiaotong University, Chengdu, 611756, China. E-mail: goods@my.swjtu.edu.cn.}
\thanks{Z. Zhou and A. R. Adhikary are with the School of Mathematics, Southwest Jiaotong University,
Chengdu, 611756, China. E-mail: zzc@swjtu.edu.cn, Avik.Adhikary@ieee.org.}
\thanks{Y. Feng is with the College of Systems Engineering, National University of Defense Technology,
Changsha, 410073, China. E-mail: fengyanghe@nudt.edu.cn}
\thanks{Pingzhi Fan is with the Institute of Mobile Communications, Southwest Jiaotong University,
	Chengdu, 611756, China. E-mail: pzfan@swjtu.edu.cn}
}
\maketitle

\date{}

\begin{abstract}
Zero correlation zone (ZCZ) sequences  and
Golay sequences are two kinds of sequences with different preferable correlation properties.
It was shown by Gong \textit{et al.}  and Chen \textit{et al.} that some Golay sequences also possess a large ZCZ and are good candidates
for pilots in OFDM systems.
Known Golay sequences with ZCZ reported in the literature have a limitation in the length which is the form of a power of 2. One objective of this paper is to propose a construction of Golay complementary pairs (GCPs) with new lengths whose periodic autocorrelation of each of the Golay sequences and periodic corss-correlation of the pair displays a zero correlation zone (ZCZ) around the in-phase position. Specifically, the proposed  GCPs have length $4N$ (where, $N$ is the length of a GCP) and ZCZ width $N+1$. Another objective of this paper is to extend the construction to two-dimensional Golay complementary array pairs (GCAPs). Interestingly the periodic corss-correlation of the proposed GACPs also have large ZCZs around the in-phase position.
\end{abstract}

\begin{IEEEkeywords}
Aperiodic correlation, binary sequence, Golay complementary sequence, Golay complementary array, zero correlation zone.
\end{IEEEkeywords}

\section{Introduction}\label{section 1}

M. J. Golay introduced Golay complementary pairs (GCPs) in his work on multislit spectrometry \cite{Golay51}. GCPs are sequence pairs having zero aperiodic autocorrelation sums (AACS) at all non-zero time shifts \cite{Golay61}. Due to their ideal correlation sums, GCPs have found numerous applications in modern day communication systems \cite{Davis1999,Paterson2000,Georghiades2001,farrel2003,abdi2007,lei2014}, Radar \cite{Spano1996,Pezeshki2008}, etc. One of the main drawbacks of the GCPs are its availability for limited lengths \cite{Borwein2000}. To overcome this drawback and to find the sequence pairs which depicts ``closest" autocorrelation properties to that of the GCPs Fan \textit{et al.} proposed Z-complementary pairs (ZCPs) in 2007 \cite{Fan2007}. ZCPs are sequence pairs having zero AACS within a certain time-shift around the in-phase position \cite{Fan2007}. In recent years lot of research has been done on the existence \cite{Li2011}, systematic constructions \cite{liu20141,liu2014,Adhikary2016,Adhikary2018,Adhikary2020,chen2017,Adhikary20201} and applications of ZCPs \cite{Adhikary20191,chen20192}.

\subsection{Sequences pairs with zero periodic crosscorrelation zone}
Since the autocorrelation of the sequence pair sum up to zero at all non-zero time-shifts (or time-shifts within a certain region in case of ZCPs) Golay sequences have been widely used to reduce peak-to-mean envelope power ratio in orthogonal frequency division multiplexing systems \cite{Davis1999,Paterson2000}. However, the sequences own periodic autocorrelation plays an important role in some applications like synchronization and detection of the signal. Working in this direction, Gong \textit{et al.} \cite{Gong2013} investigated the periodic autocorrelation behaviour of a single Golay sequence in 2013. To be more specific, Gong \textit{et al.} presented two constructions of Golay sequences of length $2^m$ each displaying a periodic zero autocorrelation zone (ZACZ) of $2^{m-2}$, and $2^{m-3}$, respectively, around the in-phase position \cite{Gong2013}. In \cite{Gong2013}, the authors also discussed the application of Golay sequences with large ZACZ for ISI channel estimation. Using Golay sequences with large ZACZ as channel estimation sequences (CES), the authors analysed the performance of
Golay-sequence-aided channel estimation in terms of the error
variance and the classical Cramer-Rao lower bound (CRLB). The performance was also compared with the well known sequences (Frank-Zadoff-Chu sequences and $m$-sequences) which are generally used for ISI channel estimation. It was shown in \cite{Gong2013} that when the channel impulse response (CIR) is within the ZACZ width then the variance of the Golay sequences attains the CRLB.

Inspired by the work of Gong \textit{et al.} \cite{Gong2013}, Chen \textit{et al.} studied the zero cross-correlation zone among the Golay sequences in 2018 and proposed Golay-ZCZ sequence sets \cite{Chen201811}. Golay-ZCZ sequence sets are sequence sets having periodic ZACZ for each sequences, periodic zero cross-correlation zone (ZCCZ) for any two sequences and also the aperiodic autocorrelation sum is zero for all non-zero time shifts. Specifically, Chen \textit{et al.} gave a systematic construction of Golay-ZCZ sequence set consisting $2^k$ sequences, each of length $2^m$ and $\min\{ZACZ,ZCCZ\}$ is $2^{m-k-1}$ \cite{Chen201811}. However, the lengths of the GCPs with large ZCZs discussed in the works of Gong \textit{et al.} and Chen \textit{et al.} are all in the powers of two \cite{Chen201811}. To the best of the authors knowledge, the problem of investigating the individual periodic autocorrelations of the GCPs and the periodic cross-correlations of the pairs when the length of the GCPs are non-power-of-two, remains largely open. An overview of of the previous works, which considers the periodic ZACZ of the individual sequences and ZCCZ of a GCP, is given in Table \ref{Table duibi}.

\begin{table}
  \centering
  \caption{Golay sequences with periodic ZACZ and ZCCZ.}
  \label{Table duibi}
\begin{tabular}{|c|c|c|c|c|}
  \hline
  Ref. & Length $N$ & Complementary sets size $M$ & ZACZ width & ZCCZ width  \\\hline

  \hline

  \cite{Gong2013} & $2^m$ & $2$ & $2^{m-2}$ or $2^{m-3}$ & Not discussed  \\
  \hline

  \cite{Chen201811} & $2^m$ & $2^k$ & $2^{m-k-1}$ & $2^{m-k-1}$ \\
  \hline

  Theorem 1 & $4N$ & $2$ & $N$ & $N$ \\
  \hline
\end{tabular}
\end{table}

Based on the discussion on using Golay sequences as CES for ISI channel estimation in \cite{Gong2013} it can be realised that our proposed constructions will add flexibility in choosing the Golay sequences of various lengths for using it as CES. Since,
in practical scenarios, a longer training sequence will give
rise to a higher training overhead, therefore, selection of the
training length is a trade-off between channel estimation performance
and training overhead. For example, let us consider
that in a practical scenario, the CIR is a length 10 vector. If only
the Golay sequences in \cite{Gong2013} or \cite{Chen201811} are considered, then one have to use a length 64 GCP which have a ZACZ width of 16. However, by our proposed constructions, a length 40 GCP which have a ZACZ width of 10 can be used as a CES as described in \cite{Gong2013}. This will improve the system performance.

\subsection{Two-dimensional complementary array pairs}
In 1978, Ohyama \textit{et al.} introduced two-dimensional (2D) sequence sets with zero side lobes in their work on image processing \cite{Ohyama1978}. In 1985, H. D. Luke presented some iterative constructions of higher-dimensional complementary codes \cite{Luke1985}. In 1990, Bomer \textit{et al.} proposed perfect binary arrays \cite{bomer1990}. In search of one-dimensional sequences with low autocorrelation magnitudes at non-zero time-shifts, in 2007, Jedwab and Parker made a remarkable progress by generalising the problem to multiple dimensions \cite{Jedwab20071}. Instead of searching sequences with low correlation properties in one dimension, the authors analysed the possibility of existence of such sequence arrays in multiple dimensions in the hope of a larger existence patterns. In \cite{Jedwab20071} the authors presented a systematic construction of $m$ dimensional Golay complementary array pairs (GCAPs) from an $m+1$ dimensional GCAPs. 2D- GCAPs are array pairs having the 2D autocorrelations of the constituent arrays sum up to zero for all non-zero time-shifts.

Generalizing the concept of ZCZ in 2D, Fan \textit{et. al} introduced binary array set with ZCZ in 2001 \cite{Tang2001}. In 2002 Hayashi proposed a class of 2D- binary sequences with ZCZ \cite{Hayashi2004}. In 2010 Cheng \textit{et al.} proposed another new class of class of 2D- binary sequences with ZCZ \cite{Cheng2010}. Recently in 2019 Chen \textit{et al.} proposed a systematic construction of 2D- Z-complementary array pairs (ZCAPs) \cite{pai2019}. However, the behavior of autocorrelation
of a single Golay array is still unknown. To the best of the authors knowledge, the problem of investigating the periodic 2D- autocorrelations of the constituent arrays of the 2D- GCAPs and the periodic 2D- cross-correlations of the 2D- GCAPs, remains largely open.

\subsection{Contributions}

Motivated by the works of Gong \textit{et al.} \cite{Gong2013} and Chen \textit{et al.} \cite{Chen201811} we propose a systematic construction of GCPs of length non-power-of-two, where the individual sequences have a periodic ZACZ and the periodic cross-correlation of the sequence pairs also have a ZCCZ. We also extend the ideas to construct 2D-GCAPs.  To be more specific we make  the following contributions in this paper:
\begin{enumerate}
	\item Assuming a GCP of length $N$ exists, we  systematically construct GCPs of length $4N$. The proposed GCPs have $Z_{\min}=N+1$, where $Z_{\min}=\min\{ZACZ,ZCCZ\}$.
	\item We  also systematically constructe a 2D GCAP of size $s_1\times 4s_2$, assuming a 2D- GCAP of size $s_1 \times s_2$ exists. The designed 2D- GCAPs have a $2D\text{-}Z_{\min}=s_1\times (s_2+1)$, where $2D\text{-}Z_{\min}=\min\{2D\text{-}ZACZ,2D\text{-}ZCCZ\}$.
	\item We propose a systematic construction of 2D GCAP of size $4s_1\times 4s_2$, assuming a 2D- GCAP of size $s_1 \times s_2$ exists. The designed 2D- GCAPs have a $2D\text{-}Z_{\min}=(s_1+1)\times (s_2+1)$, where $2D\text{-}Z_{\min}=\min\{2D\text{-}ZACZ,2D\text{-}ZCCZ\}$.
\end{enumerate}

\subsection{Organization}
The rest of the paper is organized as follows. In Section \ref{section 2}, some useful notations and preliminaries are recalled.
In Section \ref{section 3}, a systematic construction for GCPs of lengths non-power-of-two with large periodic ZACZ and ZCCZ is proposed.
In Section \ref{section 4}, we extended the construction to 2D- GCAPs.
Finally, we conclude the paper in Section \ref{section 5}.

\section{Preliminaries}\label{section 2}
\begin{definition}
	Let $\mathcal{A}=[A_{i,j}]$ and $\mathcal{B}=[B_{i,j}]$, for $0\leq i <L_1$ and $0\leq j <L_2$, be two unimodular complex-valued 2D- arrays, each of size $L_1\times L_2$. Then, for $-L_1<\tau_1<L_1$ and $-L_2<\tau_2<L_2$, the 2D- periodic crosscorrelation function is defined as follows:
	\begin{equation}
      R_{\mathcal{A},\mathcal{B}}(\tau_1,\tau_2)=\sum\limits_{i=0}^{L_1-1}\sum\limits_{j=0}^{L_2-1}A_{i,j}B^*_{(i+\tau_1)\pmod {L_1},(j+\tau_2)\pmod {L_2}},
	\end{equation}
	where $(^*)$ denotes the complex conjugate of $B_{(i+\tau_1)\pmod {L_1},(j+\tau_2)\pmod {L_2}}$. When $L_1=1$, then the above definition reduces to a 1D- periodic cross-correlation function. When $\mathcal{A}=\mathcal{B}$, it is called periodic autocorrelation function, and is denoted by $R_{\mathcal{A}}$ for short.
\end{definition}

\begin{definition}
	For $0\leq i <L_1$ and $0\leq j <L_2$, let $\mathcal{A}=[A_{i,j}]$ and $\mathcal{B}=[B_{i,j}]$ be two unimodular complex-valued 2D- arrays, each of size $L_1\times L_2$. Then the 2D- aperiodic crosscorrelation function is defined as follows:
	\begin{equation}
	C_{\mathcal{A},\mathcal{B}}(\tau_1,\tau_2)=\begin{cases}
	\sum\limits_{i=0}^{L_1-1-\tau_1}\sum\limits_{j=0}^{L_2-1-\tau_2}A_{i,j}B^*_{(i+\tau_1),(j+\tau_2)}, & 0\leq \tau_1<L_1,0\leq \tau_2<L_2;\\
	\sum\limits_{i=0}^{L_1-1+\tau_1}\sum\limits_{j=0}^{L_2-1-\tau_2}A_{i-\tau_1,j}B^*_{i,(j+\tau_2)}, & -L_1< \tau_1<0,0\leq \tau_2<L_2;\\
	\sum\limits_{i=0}^{L_1-1-\tau_1}\sum\limits_{j=0}^{L_2-1+\tau_2}A_{i,j-\tau_2}B^*_{(i+\tau_1),j}, & 0\leq \tau_1<L_1,-L_2< \tau_2<0;\\
	\sum\limits_{i=0}^{L_1-1+\tau_1}\sum\limits_{j=0}^{L_2-1+\tau_2}A_{i-\tau_1,j-\tau_2}B^*_{i,j}, & -L_1< \tau_1<0,-L_2< \tau_2<0;
	\end{cases}
	\end{equation}
	where $(^*)$ denotes the complex conjugate of the element. When $L_1=1$, then the above definition reduces to a 1D- aperiodic cross-correlation function. When $\mathcal{A}=\mathcal{B}$, it is called aperiodic autocorrelation function, and is denoted by $C_{\mathcal{A}}$.
\end{definition}

Using the above definitions, we have the following property.
\begin{property}
Let $\mathcal{A}$ be a 2D- array with size $L_1\times L_2$, then for $0\leq \tau_1<L_1$ and $0\leq \tau_2<L_2$, we have
  \begin{equation}
  \begin{split}
  C_{\mathcal{A}}(\tau_{1},\tau_2) &=C_{\mathcal{A}}^*(-\tau_{1},-\tau_2);\\
  R_{\mathcal{A}}(\tau_{1},\tau_2) &=R_{\mathcal{A}}^*(-\tau_{1},-\tau_2).
  \end{split}
  \end{equation}
\end{property}

\begin{definition}\label{Definition-GCAP}
A pair of 2D- arrays $\mathcal{A}$ and $\mathcal{B}$, each of size $L_1\times L_2$ is called a Golay complementary array pair (GCAP) if
\begin{equation}
  C_\mathcal{A}(\tau_1,\tau_2)+C_\mathcal{B}(\tau_1,\tau_2)=0, \hbox{ for all } (\tau_1,\tau_2)\neq(0,0).
\end{equation}
In particular, when $L_1=1$, GCAP is reduced to a Golay complementary pair (GCP), denoted by $(\mathbf{a,b})$.
\end{definition}

\begin{definition}\label{Definition-GAMate}
Let $(\mathcal{A,B})$ and $(\mathcal{C,D})$ be two GCAPs, each of size $L_1\times L_2$. Then
$(\mathcal{C,D})$ is called a Golay array mate of $(\mathcal{A,B})$ if \cite{Matsufuji2010}
$$C_\mathcal{A,C}(\tau_1,\tau_2)+C_\mathcal{B,D}(\tau_1,\tau_2)=0,$$
for any $-L_1<\tau_1<L_1$ and $-L_2<\tau_2<L_2$.
In particular, when $L_1=1$, $(\mathcal{A,B})$ reduces to a GCP and $(\mathcal{C,D})$ reduces to mate of the GCP $(\mathcal{A,B})$.
\end{definition}

\begin{lemma}
Let $(\mathcal{A,B})$ be GCAP of size $L_1\times L_2$. Then one of the mate of $(\mathcal{A,B})$ is as follows:
\begin{equation}
	(\mathcal{C,D}) =(\overleftarrow{\mathcal{B^*}},-\overleftarrow{\mathcal{A^*}}),
\end{equation}
where $\overleftarrow{\mathcal{A}}$ denote reverse of array $\mathcal{A}$ at every dimension and $(^*)$ represents complex conjugation.
\end{lemma}

%
\begin{definition}\label{Definition-GolayZCZ}
Let ($\mathcal{A}$,$\mathcal{B}$) be a GCAP of size $L_1\times L_2$. Let the 2D- periodic autocorrelation function of $\mathcal{A}$ and $\mathcal{B}$ have a 2D- ZACZ of $Z_{k_0}$ and $Z_{k_1}$, respectively. Also let the 2D- periodic cross-correlation function have a 2D- ZCCZ of $Z_{k_2}$ when the cross-correlation between $\mathcal{A}$ and $\mathcal{B}$ is considered. Then ($\mathcal{A}$,$\mathcal{B}$) is a GCAP having a 2D- periodic ZCZ of $Z_{\min}=\min\{Z_{k_0},Z_{k_1},Z_{k_2}\}$ and we denote it as
$(L_1\times L_2,Z_{\min})$- GCAP. In other words, the 2D- array pair ($\mathcal{A}$,$\mathcal{B}$) is a $(L_1\times L_2,Z_{\min})$- GCAP, having 2D- periodic ZCZ $Z_{\min}$ if
\begin{equation}
\begin{split}
&\text{C1: }C_\mathcal{A}(\tau_1,\tau_2)+C_\mathcal{B}(\tau_1,\tau_2)=0, \hbox{ for all } (\tau_1,\tau_2)\neq(0,0);\\
&\text{C2: }R_\mathcal{A}(\tau_1,\tau_2)=R_\mathcal{B}(\tau_1,\tau_2)=0, \hbox{ for } |\tau_1|<Z_{\min},|\tau_2|<Z_{\min} \text{ and }(\tau_1,\tau_2)\neq(0,0);\\
&\text{C3: }R_{\mathcal{A},\mathcal{B}}(\tau_1,\tau_2)=0, \hbox{ for } |\tau_1|<Z_{\min},|\tau_2|<Z_{\min},
\end{split}
\end{equation}
where, $Z_{\min}$ is defined as above. When $L_1=1$, we get the conditions for $(L_2,Z_{\min})$- GCPs having 1D- periodic ZCZ width of $Z_{\min}$.
\end{definition}

\section{New Construction of Golay Complementary Pair with ZCZ}\label{section 3}
In this section we propose a new class of GCPs, where each of the sequences have a periodic ZACZ and also have a periodic ZCCZ when the cross-correlation between the pairs are considered. Please note that GCPs are 1D- GCAPs where $L_1=1$ and $L_2=N$ (in line with Section II). In this section, $\mathbf{a}\|\mathbf{b}$ denotes the concatenation of sequences $\mathbf{a}$ and $\mathbf{b}$ and $Z_{\min}$ is usually 1D- $Z_{\min}$. Before introduce the construction, we need the following lemma.

\begin{lemma}\label{lem2}
	Let $(\mathcal{A,B})$ a be GCAP of size $s_1\times s_2$ and $(\mathcal{C,D})$ be one of the Golay array mate of $(\mathcal{A,B})$. Let
	\begin{equation}
	\mathcal{P}=
	\left[
	\begin{array}{rrrr}
	x_1\mathcal{A} & x_2\mathcal{B} & x_3\mathcal{A} & x_4\mathcal{B} \\
	\end{array}
	\right],
	\mathcal{Q}=
	\left[
	\begin{array}{rrrr}
	x_1\mathcal{C} & x_2\mathcal{D} & x_3\mathcal{C} & x_4\mathcal{D} \\
	\end{array}
	\right],
	\end{equation}
	where $x_1,x_2,x_3,x_4\in\{+1,-1\}$. Then $(\mathcal{P,Q})$ is a GCAP of size $s_1\times 4s_2$, if the following condition holds:
	\begin{equation}
		x_1x_2+x_3x_4=0.
	\end{equation}
\end{lemma}
\begin{IEEEproof} Let us consider $0\leq\tau_1\leq s_1-1$ and $0\leq\tau_2\leq s_2-1$. Then we have
	\begin{equation}
	\begin{split}
	C_\mathcal{P}(\tau_1,\tau_2)=&2\big(C_\mathcal{A}(\tau_1,\tau_2) +C_\mathcal{B}(\tau_1,\tau_2)\big)+(x_1x_2+x_3x_4)C_\mathcal{B,A}^*(\tau_1,s_2-\tau_2) +x_2x_3\rho_\mathcal{A,B}^*(\tau_1,s_2-\tau_2), \\
	C_\mathcal{Q}(\tau_1,\tau_2)=&2\big(C_\mathcal{C}(\tau_1,\tau_2) +C_\mathcal{D}(\tau_1,\tau_2)\big)+(x_1x_2+x_3x_4)C_\mathcal{D,C}^*(\tau_1,s_2-\tau_2) +x_2x_3\rho_\mathcal{C,D}^*(\tau_1,s_2-\tau_2).
	\end{split}
	\end{equation}
	Hence, for $0\leq\tau_1\leq s_1-1$ and $0\leq\tau_2\leq s_2-1$, we have
	\begin{equation}
		C_\mathcal{P}(\tau_1,\tau_2)+C_\mathcal{Q}(\tau_1,\tau_2)=4\big(C_\mathcal{A}(\tau_1,\tau_2) +C_\mathcal{B}(\tau_1,\tau_2)\big).
	\end{equation}

Consider $0\leq\tau_1\leq s_1-1$ and $s_2\leq\tau_2\leq 2s_2-1$. Then one has
\begin{equation}
\begin{split}
C_\mathcal{P}(\tau_1,\tau_2)=& (x_1x_2+x_3x_4)C_\mathcal{A,B}(\tau_1,\tau_2-s_2) +x_2x_3C_\mathcal{B,A}(\tau_1,\tau_2-s_2) +x_1x_3C_\mathcal{A}^*(\tau_1,2s_2-\tau_2)\\&\hspace{1in} +x_2x_4C_\mathcal{B}^*(\tau_1,2s_2-\tau_2), \\
C_\mathcal{Q}(\tau_1,\tau_2)=& (x_1x_2+x_3x_4)C_\mathcal{C,D}(\tau_1,\tau_2-s_2) +x_2x_3C_\mathcal{D,C}(\tau_1,\tau_2-s_2) +x_1x_3C_\mathcal{C}^*(\tau_1,2s_2-\tau_2)\\&\hspace{1in} +x_2x_4C_\mathcal{D}^*(\tau_1,2s_2-\tau_2).
\end{split}
\end{equation}

Hence, for $0\leq\tau_1\leq s_1-1$ and $s_2\leq\tau_2\leq 2s_2-1$, we have
\begin{equation}
C_\mathcal{P}(\tau_1,\tau_2)+C_\mathcal{Q}(\tau_1,\tau_2)=0.
\end{equation}

Consider $0\leq\tau_1\leq s_1-1$ and $2s_2\leq\tau_2\leq 3s_2-1$. Then we have
\begin{equation}
\begin{split}
C_\mathcal{P}(\tau_1,\tau_2)=& x_1x_3C_\mathcal{A}(\tau_1,\tau_2-2s_2) +x_2x_4C_\mathcal{B}(\tau_1,\tau_2-2s_2) +x_1x_4C_\mathcal{B,A}^*(\tau_1,3s_2-\tau_2), \\
C_\mathcal{Q}(\tau_1,\tau_2)=& x_1x_3C_\mathcal{C}(\tau_1,\tau_2-2s_2) +x_2x_4C_\mathcal{D}(\tau_1,\tau_2-2s_2) +x_1x_4C_\mathcal{D,C}^*(\tau_1,3s_2-\tau_2).
\end{split}
\end{equation}

Hence, for $0\leq\tau_1\leq s_1-1$ and $2s_2\leq\tau_2\leq 3s_2-1$, we have
\begin{equation}
C_\mathcal{P}(\tau_1,\tau_2)+C_\mathcal{Q}(\tau_1,\tau_2)=0.
\end{equation}

Consider $0\leq\tau_1\leq s_1-1$ and $3s_2\leq\tau_2\leq 4s_2-1$. Then we have
\begin{equation}
\begin{split}
C_\mathcal{P}(\tau_1,\tau_2)=& x_1x_4C_\mathcal{A,B}(\tau_1,\tau_2-3s_2), \\
C_\mathcal{Q}(\tau_1,\tau_2)=& x_1x_4C_\mathcal{C,D}(\tau_1,\tau_2-3s_2),.
\end{split}
\end{equation}

Hence, for $0\leq\tau_1\leq s_1-1$ and $3s_2\leq\tau_2\leq 4s_2-1$, we have
\begin{equation}
C_\mathcal{P}(\tau_1,\tau_2)+C_\mathcal{Q}(\tau_1,\tau_2)=0.
\end{equation}

The other cases can also be proved similarly.
\end{IEEEproof}

\begin{construction}\label{Construction-seq-ZCZ}
Let $(\mathbf{a,b})$ be a GCP of length $N$ and $(\mathbf{c,d})$ be the Golay mate of $(\mathbf{a,b})$, then define
\begin{equation}
  \begin{split}
    \mathbf{p}=~&x_1\cdot\mathbf{a}~\|~x_2\cdot\mathbf{b}~\|~x_3\cdot\mathbf{a}~\|x_4\cdot\mathbf{b}, \\
    \mathbf{q}=~&x_1\cdot\mathbf{c}~\|~x_2\cdot\mathbf{d}~\|~x_3\cdot\mathbf{c}~\|x_4\cdot\mathbf{d},
  \end{split}
\end{equation}
where $x_1,x_2,x_3,x_4\in\{+1,-1\}$.
\end{construction}

\begin{theorem}
  Let $(\mathbf{p,q})$ be a sequence pair generated via Construction \ref{Construction-seq-ZCZ}. Then $(\mathbf{p,q})$ is a GCP of length $4N$ with $Z_{\min}=N+1$, if $x_1,x_2,x_3,x_4$ if the following condition holds:
  \begin{equation}
  	x_1x_2+x_3x_4=0.
  \end{equation}
\end{theorem}
\begin{IEEEproof}
Using Lemma \ref{lem2} for 1D cases, i.e., considering $s_1=1$ and $s_2=N$, we can prove that $(\mathbf{p,q})$ is a GCP of length $4N$.	

Next, we have to prove the other two conditions of Definition \ref{Definition-GolayZCZ}. Consider $1\leq\tau\leq N$, then we have
\begin{equation}
\begin{split}
R_\mathbf{p}(\tau)=& \sum_{k=0}^{4N-1} p_kp_{k+\tau}^*\\
=& \sum_{k=0}^{N-1-\tau} a_ka_{k+\tau}^* +\sum_{k=N-\tau}^{N-1} x_1a_kx_2^*b_{k-(N-\tau)}^* +\sum_{k=0}^{N-1-\tau} b_kb_{k+\tau}^* +\sum_{k=N-\tau}^{N-1} x_2b_kx_3^*a_{k-(N-\tau)}^* +\\
& \sum_{k=0}^{N-1-\tau} a_ka_{k+\tau}^* +\sum_{k=N-\tau}^{N-1} x_3a_kx_4^*b_{k-(N-\tau)}^* +\sum_{k=0}^{N-1-\tau} b_kb_{k+\tau}^* +\sum_{k=N-\tau}^{N-1} x_4b_kx_1^*a_{k-(N-\tau)}^* \\
=&2\big(\rho_\mathbf{a}(\tau) +\rho_\mathbf{b}(\tau)\big). \\
\end{split}
\end{equation}
Similarly for $1\leq\tau\leq N$, we have
\begin{equation}
\begin{split}
R_\mathbf{q}(\tau)=&2\big(C_\mathbf{c}(\tau) +C_\mathbf{d}(\tau)\big), \\
R_\mathbf{p,q}(\tau)=&2\big(C_\mathbf{a,c}(\tau) +C_\mathbf{b,d}(\tau)\big).
\end{split}
\end{equation}
Since $(\mathbf{a},\mathbf{b})$ is a GCP and $(\mathbf{c},\mathbf{d})$ is one of the complementary mates of $(\mathbf{a},\mathbf{b})$, $R_\mathbf{p}(\tau)=R_\mathbf{q}(\tau)=0,\hbox{ for all } 1\leq\tau\leq N,$ and $R_\mathbf{p,q}(\tau)=0,\hbox{ for all } 0\leq\tau\leq N$.

Therefore, $(\mathbf{p,q})$ is a GCP of length $4N$ with $Z_{\min}=N+1$.
\end{IEEEproof}


\begin{example}\label{ex1 binary}
Let $(\mathbf{a,b})$ be a binary GCP of length 10, given by $\mathbf{a}=(1,1,-1,1,1,1,1,1,-1,-1),\mathbf{b}=(1,1,-1,1,-1,1,-1, -1,1,1)$.
Then $(\mathbf{c,d})=(\overleftarrow{\mathbf{b}^*},-\overleftarrow{\mathbf{a}^*})$ is a Golay mate of $(\mathbf{a,b})$. Define
\begin{equation}
  \begin{split}
    \mathbf{p}=~&\mathbf{a}~\|~\mathbf{b}~\|~\mathbf{a}~\|-\mathbf{b}, \\
    \mathbf{q}=~&\mathbf{c}~\|~\mathbf{d}~\|~\mathbf{c}~\|-\mathbf{d}.
  \end{split}
\end{equation}
Then $(\mathbf{p,q})$ is a GCP of length $40$ with $Z_{\min}=11$, because
\begin{equation}
  \begin{split}
    \big(R_\mathbf{p}(\tau)\big)_{\tau=0}^{39}=&(40,\mathbf{0}_{10},-4,-8,4,8, -4,0,4,0,12,0,12,0,4,0,-4,8,4,-8,-4,\mathbf{0}_{10}), \\
    \big(R_\mathbf{q}(\tau)\big)_{\tau=0}^{39}=&(40,\mathbf{0}_{10},4,8,-4,-8, 4,0,-4,0,-12,0,-12,0,-4,0,4,-8,-4,8,4,\mathbf{0}_{10}), \\
    \big(R_\mathbf{p,q}(\tau)\big)_{\tau=0}^{39}=&(\mathbf{0}_{11},-4,-8,4, 16,4,0,4,-8,-4,0,4,-8,12,0,12,0,-4,8,4,\mathbf{0}_{10}).
  \end{split}
\end{equation}
\end{example}

\begin{example}\label{ex1}
Let $\mathbf{a}=(1,1,-1),\mathbf{b}=(1,i,1)$, then $(\mathbf{a,b})$ be a quadriphase GCP of length 3.
Then $(\mathbf{c,d})=(\overleftarrow{\mathbf{b}^*},-\overleftarrow{\mathbf{a}^*})$ is a Golay mate of $(\mathbf{a,b})$. Define
\begin{equation}
  \begin{split}
    \mathbf{p}=~&\mathbf{a}~\|~\mathbf{b}~\|~\mathbf{a}~\|-\mathbf{b} =(1,1,-1,1,i,1,1,1,-1,-1,-i,-1), \\
    \mathbf{q}=~&\mathbf{c}~\|~\mathbf{d}~\|~\mathbf{c}~\|-\mathbf{d} =(1,-i,1,1,-1,-1,1,-i,1,-1,1,1).
  \end{split}
\end{equation}
Then $(\mathbf{p,q})$ is a GCP of length $12$ with $Z_{\min}=4$, because
\begin{equation}
  \begin{split}
    \big(R_\mathbf{p}(\tau)\big)_{\tau=0}^{11}=&(12,0,0,0,-4,0,0,0,-4,0,0,0), \\
    \big(R_\mathbf{q}(\tau)\big)_{\tau=0}^{11}=&(12,0,0,0,4,0,0,0,4,0,0,0), \\
    \big(R_\mathbf{p,q}(\tau)\big)_{\tau=0}^{11}=&(0,0,0,0,-4,4-4i,4i,4+4i,4,0,0,0).
  \end{split}
\end{equation}
The periodic correlation magnitudes of $(12,4)$- GCP ($\mathbf{p,q}$) are shown in Fig. \ref{fig1}.
\begin{figure}[ht]
  \centering
  \includegraphics[width=.7\textwidth]{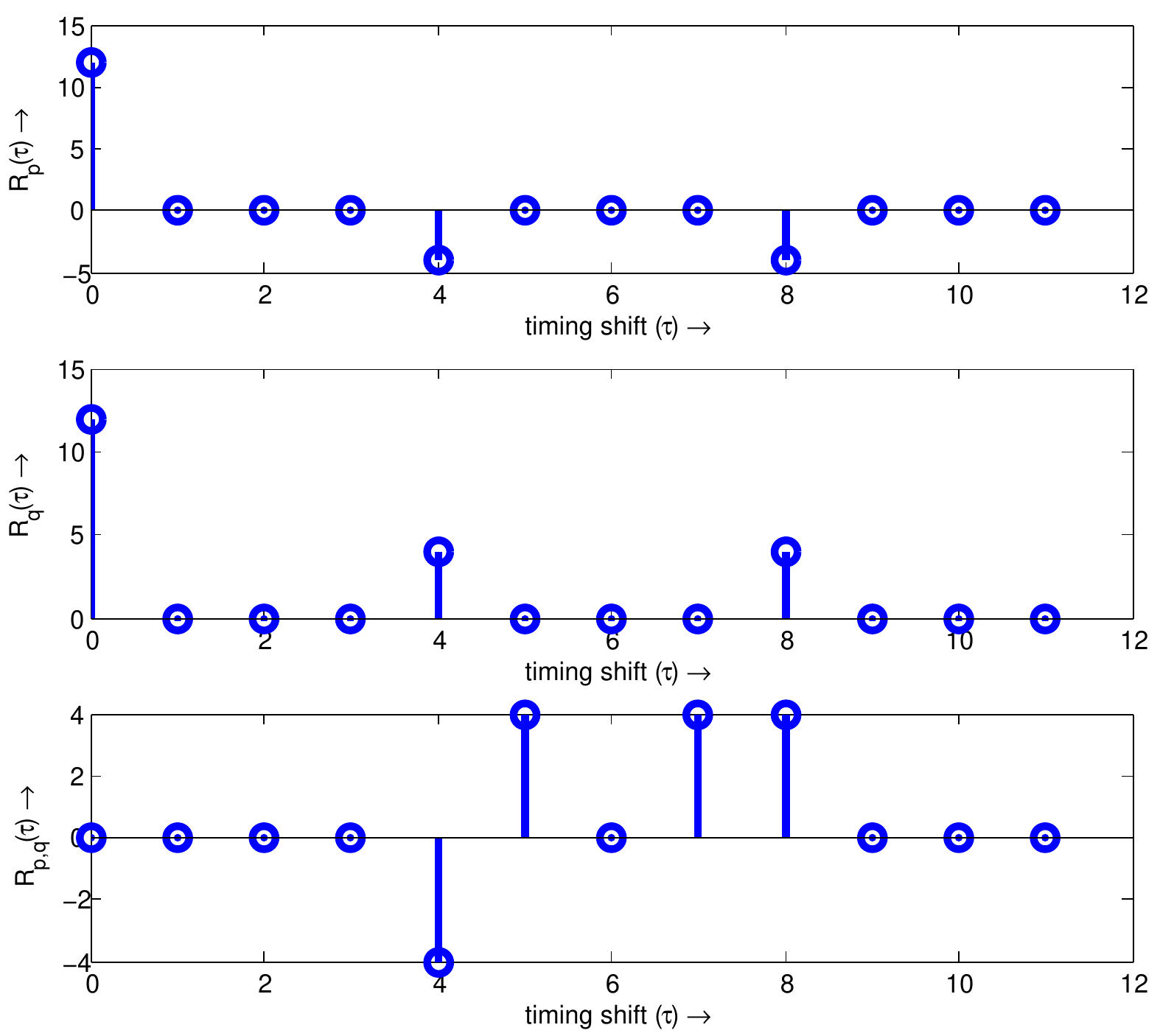}
  \caption{A glimpse of the 1D- periodic correlations of the proposed GCP given in Example \ref{ex1}.}\label{fig1}
\end{figure}
\end{example}

\begin{remark}
	Please note that binary GCP of length 40 and quadriphase GCP of length 12 with ZCZ has not been previously reported in the literature. By taking GCP $(\mathbf{a,b})$ of length $2^m$, then length of $(\mathbf{p,q})$ is $2^{m+2}$ and the generated GCP $(\mathbf{p,q})$ is reduced to the GCP in \cite{Gong2013}.
\end{remark}

\section{New Construction of Golay Complementary Array with ZCZ}\label{section 4}
In this section we have proposed new sets of GCAPs, where each of the arrays have a 2D- periodic ZACZ and also have a 2D- periodic ZCCZ when the cross-correlation between the array pairs are considered. In this section $Z_{\min}$ are usually 2D- $Z_{\min}$.
\begin{construction}\label{Construction-array-ZCZ1}
Let $(\mathcal{A,B})$ be a GCAP of size $s_1\times s_2$ and $(\mathcal{C,D})$ be the Golay array mate of $(\mathcal{A,B})$, then define
\begin{equation}
  \mathcal{P}=
  \left[
    \begin{array}{rrrr}
      x_1\mathcal{A} & x_2\mathcal{B} & x_3\mathcal{A} & x_4\mathcal{B} \\
    \end{array}
  \right],
  \mathcal{Q}=
  \left[
    \begin{array}{rrrr}
      x_1\mathcal{C} & x_2\mathcal{D} & x_3\mathcal{C} & x_4\mathcal{D} \\
    \end{array}
  \right],
\end{equation}
where $x_1,x_2,x_3,x_4\in\{+1,-1\}$.
\end{construction}

\begin{theorem}\label{Theorem-array-ZCZ1}
  Let $(\mathcal{P},\mathcal{Q})$ be an array pair constructed via Construction \ref{Construction-array-ZCZ1}. Then $(\mathcal{P},\mathcal{Q})$ is a GCAP of size $s_1\times 4s_2$ with $Z_{\min}=s_1\times (s_2+1)$, if $x_1,x_2,x_3,x_4$ if the following condition holds:
  \begin{equation}
  	x_1x_2+x_3x_4=0.
  \end{equation}
\end{theorem}
\begin{IEEEproof}
	Using Lemma \ref{lem2}, we can show that $(\mathcal{P,Q})$ is a GCAP of size $s_1\times4s_2$.

Next, we have to prove the other two conditions of Definition \ref{Definition-GolayZCZ}. Consider	$-(s_1-1)\leq\tau_1\leq s_1-1,-(s_2-1)\leq\tau_2\leq (s_2-1)$. Then we have
\begin{equation}
\begin{split}
R_\mathcal{P}(\tau_1,\tau_2)=&2\big(C_\mathcal{A}(\tau_1,\tau_2) +C_\mathcal{B}(\tau_1,\tau_2) +C_\mathcal{A}(\tau_1-s_1,\tau_2) +C_\mathcal{B}(\tau_1-s_1,\tau_2)\big), \\
R_\mathcal{Q}(\tau_1,\tau_2)=&2\big(C_\mathcal{C}(\tau_1,\tau_2) +C_\mathcal{D}(\tau_1,\tau_2) +C_\mathcal{C}(\tau_1-s_1,\tau_2) +C_\mathcal{D}(\tau_1-s_1,\tau_2)\big), \\
R_\mathcal{P,Q}(\tau_1,\tau_2)=&2\big(C_\mathcal{A,C}(\tau_1,\tau_2) +C_\mathcal{B,D}(\tau_1,\tau_2) +C_\mathcal{A,C}(\tau_1-s_1,\tau_2) +C_\mathcal{B,D}(\tau_1-s_1,\tau_2)\big),
\end{split}
\end{equation}
Since $(\mathcal{A},\mathcal{B})$ is a GCAP and $(\mathcal{C},\mathcal{D})$ is one of the complementary mates of $(\mathcal{A},\mathcal{B})$, $R_\mathcal{P}(\tau_1,\tau_2)=R_\mathcal{Q}(\tau_1,\tau_2)=0,\hbox{ for all } -(s_1-1)\leq\tau_1\leq (s_1-1),-(s_2-1)\leq\tau_2\leq (s_2-1) \text{ except }(\tau_1,\tau_2)\neq(0,0),$ and $R_\mathcal{P,Q}(\tau_1,\tau_2)=0,\hbox{ for all } -(s_1-1)\leq\tau_1\leq s_1-1,-(s_2-1)\leq\tau_2\leq (s_2-1)$.

Therefore, $(\mathcal{P,Q})$ is a GCAP with periodic $Z_{\min}=s_1\times (s_2+1)$.
\end{IEEEproof}

\begin{example}\label{ex2}
  Let $\mathcal{A}=\left[\begin{array}{rrr} 1 & 1 & -1 \\ -1 & -i & -1 \\ \end{array}\right],\mathcal{B}=\left[\begin{array}{rrr} 1 & 1 & -1 \\ 1 & i & 1 \\ \end{array}\right]$, then $(\mathcal{A,B})$ is a quaternary GCAP of size $2\times3$.
  Then $(\mathcal{C,D})= (\overleftarrow{\mathcal{B^*}},-\overleftarrow{\mathcal{A^*}})$ is a Golay array mate of $(\mathcal{A,B})$. Define
\begin{equation}
  \mathcal{P}=
  \left[
    \begin{array}{rrrr}
      \mathcal{A} & \mathcal{B} & \mathcal{A} & \mathcal{-B} \\
    \end{array}
  \right],
  \mathcal{Q}=
  \left[
    \begin{array}{rrrr}
      \mathcal{C} & \mathcal{D} & \mathcal{C} & \mathcal{-D} \\
    \end{array}
  \right].
\end{equation}
Then $(\mathcal{P,Q})$ is a GCAP of size $2\times12$ with $Z_{\min}=2\times4$.
The 2D- periodic correlation magnitudes of $(2\times12,2\times4)$- GCAP $(\mathcal{P,Q})$ are shown in Fig. \ref{fig2}.
\begin{figure}[ht]
	\begin{subfigure}{.45\textwidth}
		\centering
		\includegraphics[width=\textwidth]{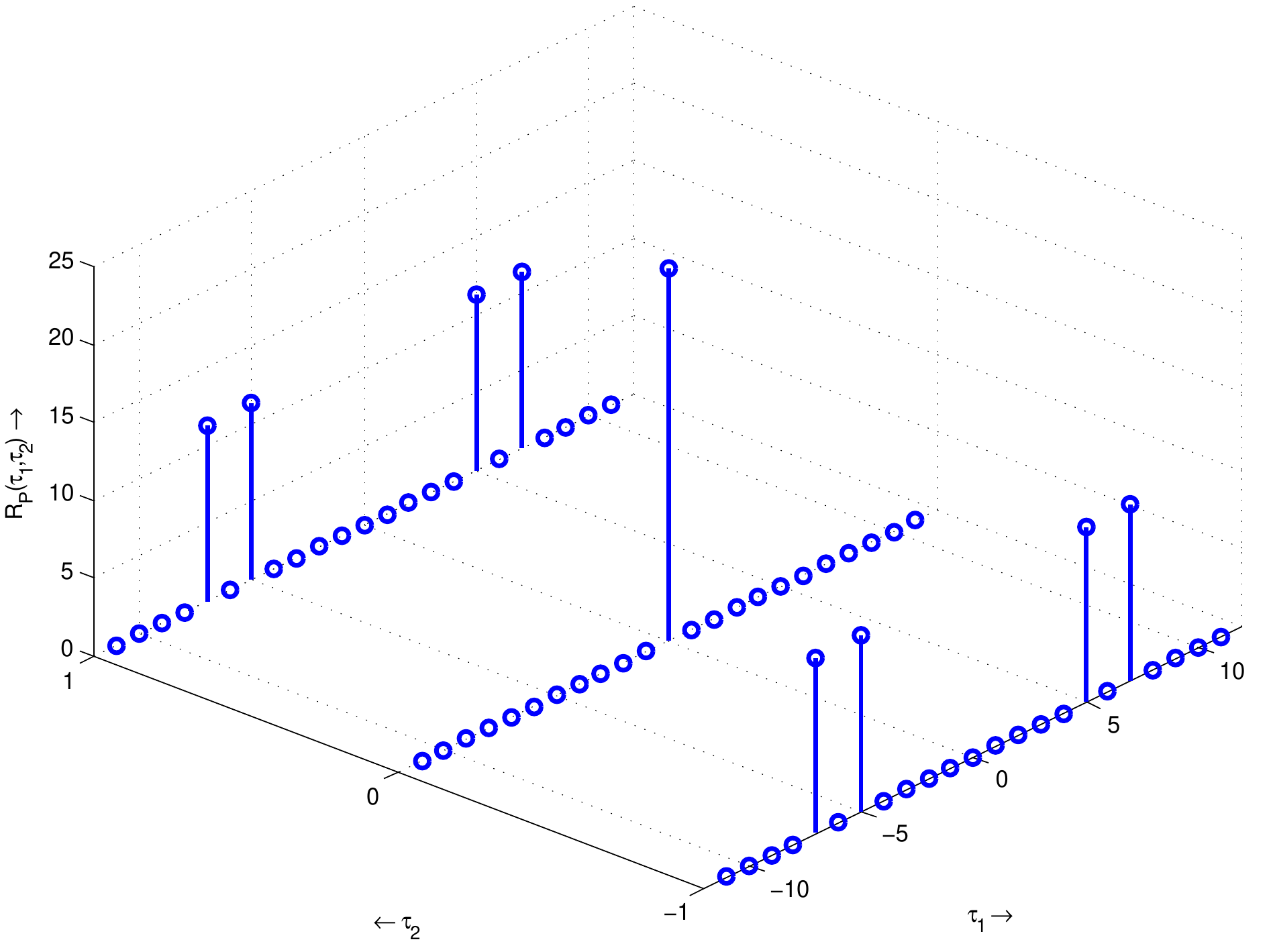}
		\caption{2D- periodic autocorrelation of $\mathcal{P}$}
		\label{fig:sub-first}
	\end{subfigure}
	\begin{subfigure}{.45\textwidth}
		\centering
		\includegraphics[width=\textwidth]{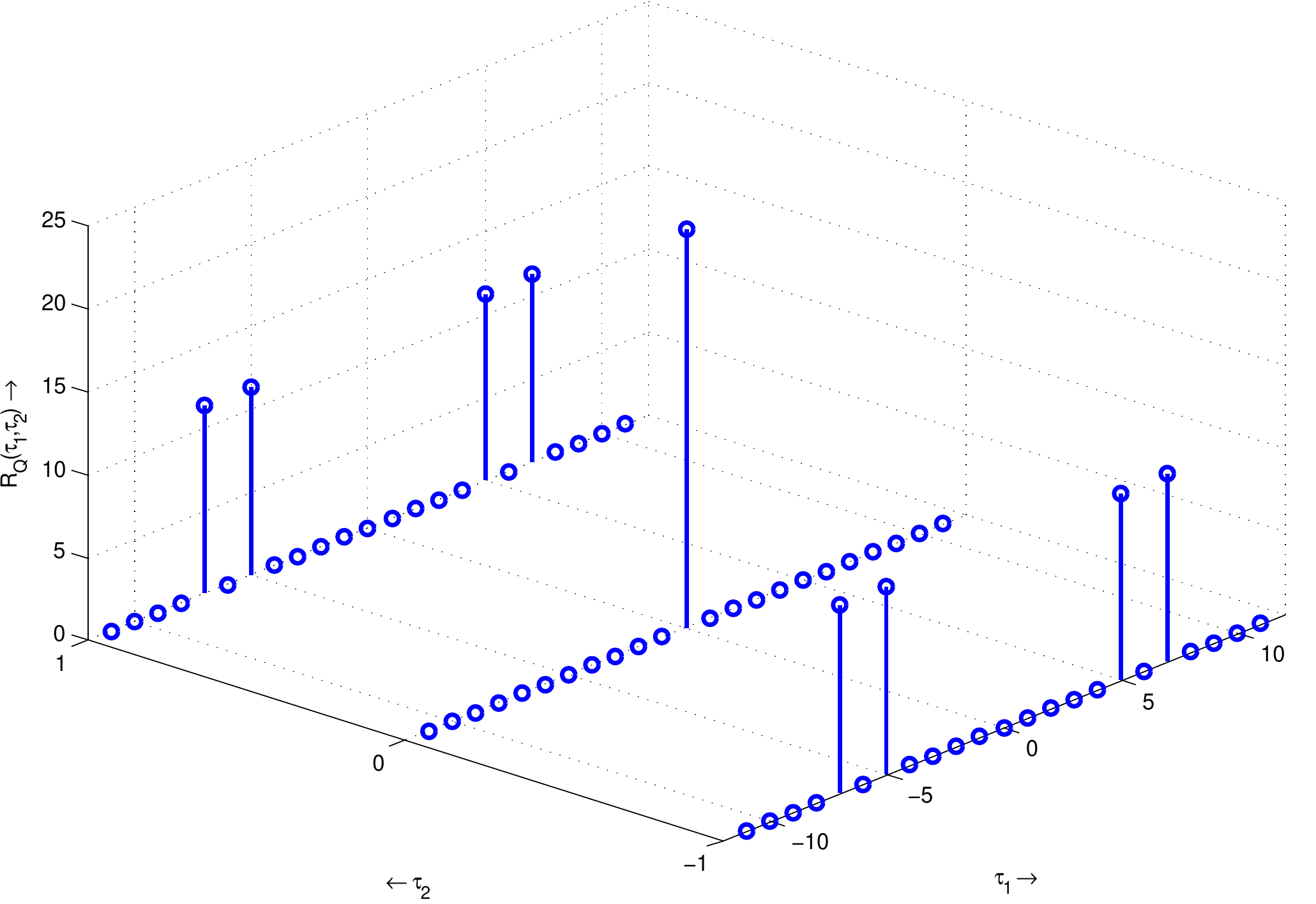}
		\caption{2D- periodic autocorrelation of $\mathcal{Q}$}
		\label{fig:sub-second}
	\end{subfigure}
\begin{subfigure}{.45\textwidth}
	\centering
	\includegraphics[width=\textwidth]{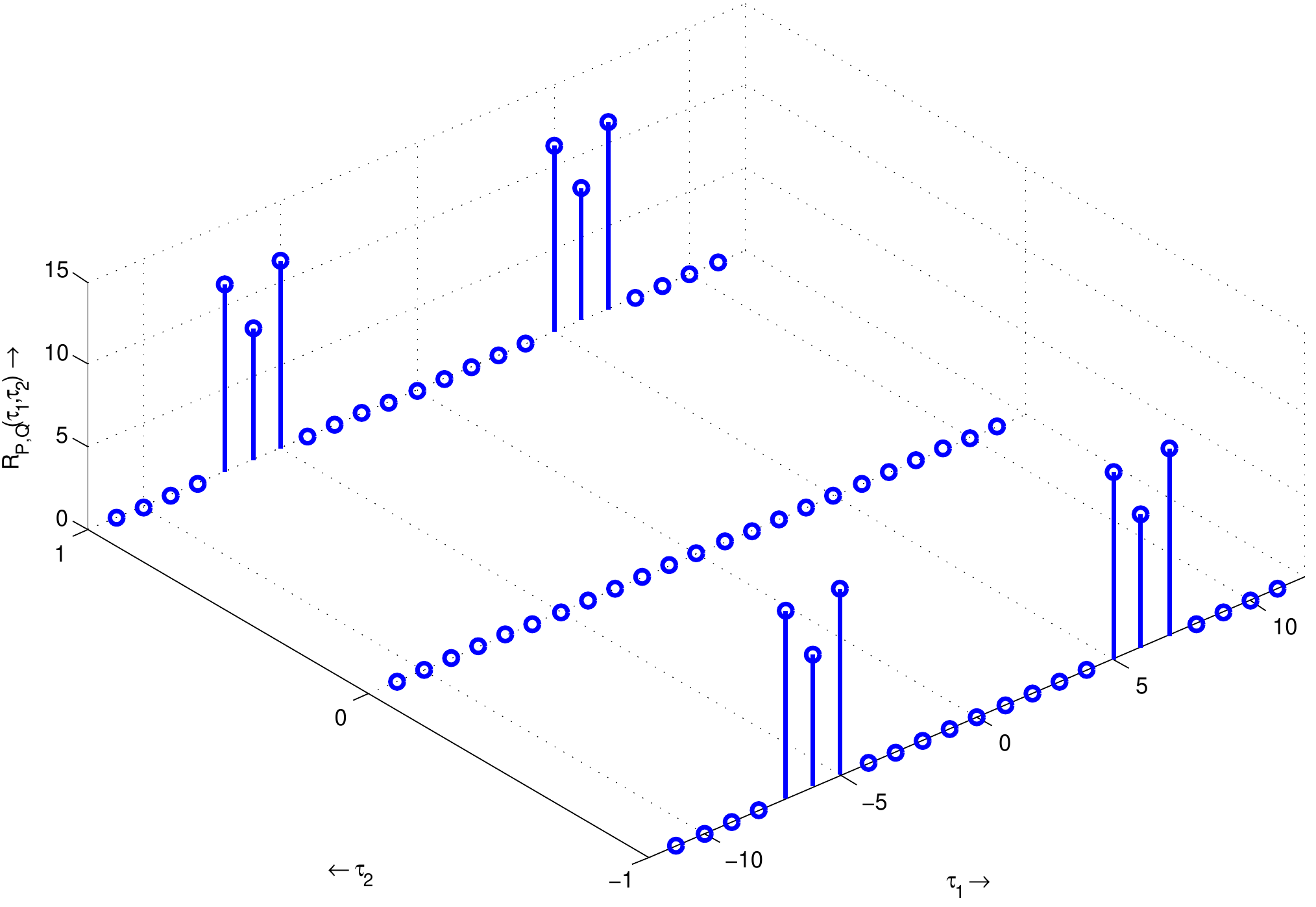}
	\caption{2D- periodic cross-correlation of $\mathcal{P}$ and $\mathcal{Q}$}
	\label{fig:sub-second}
\end{subfigure}
	\caption{A glimpse of the 2D- periodic correlations of the GCAP given in Example \ref{ex2}.}
	\label{fig2}
\end{figure}
\end{example}

\begin{construction}\label{Construction-array-ZCZ2}
Let $(\mathcal{A,B})$ be a GCAP of size $s_1\times s_2$ and $(\mathcal{C,D})$ be the Golay array mate of $(\mathcal{A,B})$, then define
\begin{equation}
  \mathcal{P}=
  \left[
    \begin{array}{rrrr}
      \mathcal{A} & \mathcal{B} & \mathcal{A} & \mathcal{-B} \\
      \mathcal{A} & \mathcal{B} & \mathcal{-A} & \mathcal{B} \\
      \mathcal{A} & \mathcal{B} & \mathcal{A} & \mathcal{-B} \\
      \mathcal{-A} & \mathcal{-B} & \mathcal{A} & \mathcal{-B} \\
    \end{array}
  \right],
  \mathcal{Q}=
  \left[
    \begin{array}{rrrr}
      \mathcal{C} & \mathcal{D} & \mathcal{C} & \mathcal{-D} \\
      \mathcal{C} & \mathcal{D} & \mathcal{-C} & \mathcal{D} \\
      \mathcal{C} & \mathcal{D} & \mathcal{C} & \mathcal{-D} \\
      \mathcal{-C} & \mathcal{-D} & \mathcal{C} & \mathcal{-D} \\
    \end{array}
  \right].
\end{equation}
\end{construction}

\begin{theorem}
  For the array pair $(\mathcal{P},\mathcal{Q})$ generated by Construction \ref{Construction-array-ZCZ2}, it's a GCAP of size $4s_1\times 4s_2$ with periodic zero correlation zone $(s_1+1)\times (s_2+1)$.
\end{theorem}
\begin{IEEEproof}
The proof is similar to the one of Theorem \ref{Theorem-array-ZCZ1}.
\end{IEEEproof}

\begin{example}\label{ex3}
  Let $\mathcal{A}=\left[\begin{array}{rrr} 1 & 1 & -1 \\ -1 & -i & -1 \\ \end{array}\right],\mathcal{B}=\left[\begin{array}{rrr} 1 & 1 & -1 \\ 1 & i & 1 \\ \end{array}\right]$, then $(\mathcal{A,B})$ is a quaternary GCAP of size $2\times3$.
  Construct $(\mathcal{C,D})= (\overleftarrow{\mathcal{B^*}},-\overleftarrow{\mathcal{A^*}})$ as a Golay array mate of $(\mathcal{A,B})$. Define
\begin{equation}
  \mathcal{P}=
  \left[
    \begin{array}{rrrr}
      \mathcal{A} & \mathcal{B} & \mathcal{A} & \mathcal{-B} \\
      \mathcal{A} & \mathcal{B} & \mathcal{-A} & \mathcal{B} \\
      \mathcal{A} & \mathcal{B} & \mathcal{A} & \mathcal{-B} \\
      \mathcal{-A} & \mathcal{-B} & \mathcal{A} & \mathcal{-B} \\
    \end{array}
  \right],
  \mathcal{Q}=
  \left[
    \begin{array}{rrrr}
      \mathcal{C} & \mathcal{D} & \mathcal{C} & \mathcal{-D} \\
      \mathcal{C} & \mathcal{D} & \mathcal{-C} & \mathcal{D} \\
      \mathcal{C} & \mathcal{D} & \mathcal{C} & \mathcal{-D} \\
      \mathcal{-C} & \mathcal{-D} & \mathcal{C} & \mathcal{-D} \\
    \end{array}
  \right].
\end{equation}
Then $(\mathcal{P,Q})$ is a GCAP of size $8\times12$ with $Z_{\min}=3\times4$.
The 2D- periodic correlation magnitudes of $(8\times12,3\times4)$- GCAP $(\mathcal{P,Q})$ are shown in Fig. \ref{fig3}.
\begin{figure}[ht]
	\begin{subfigure}{.45\textwidth}
		\centering
		\includegraphics[width=\textwidth]{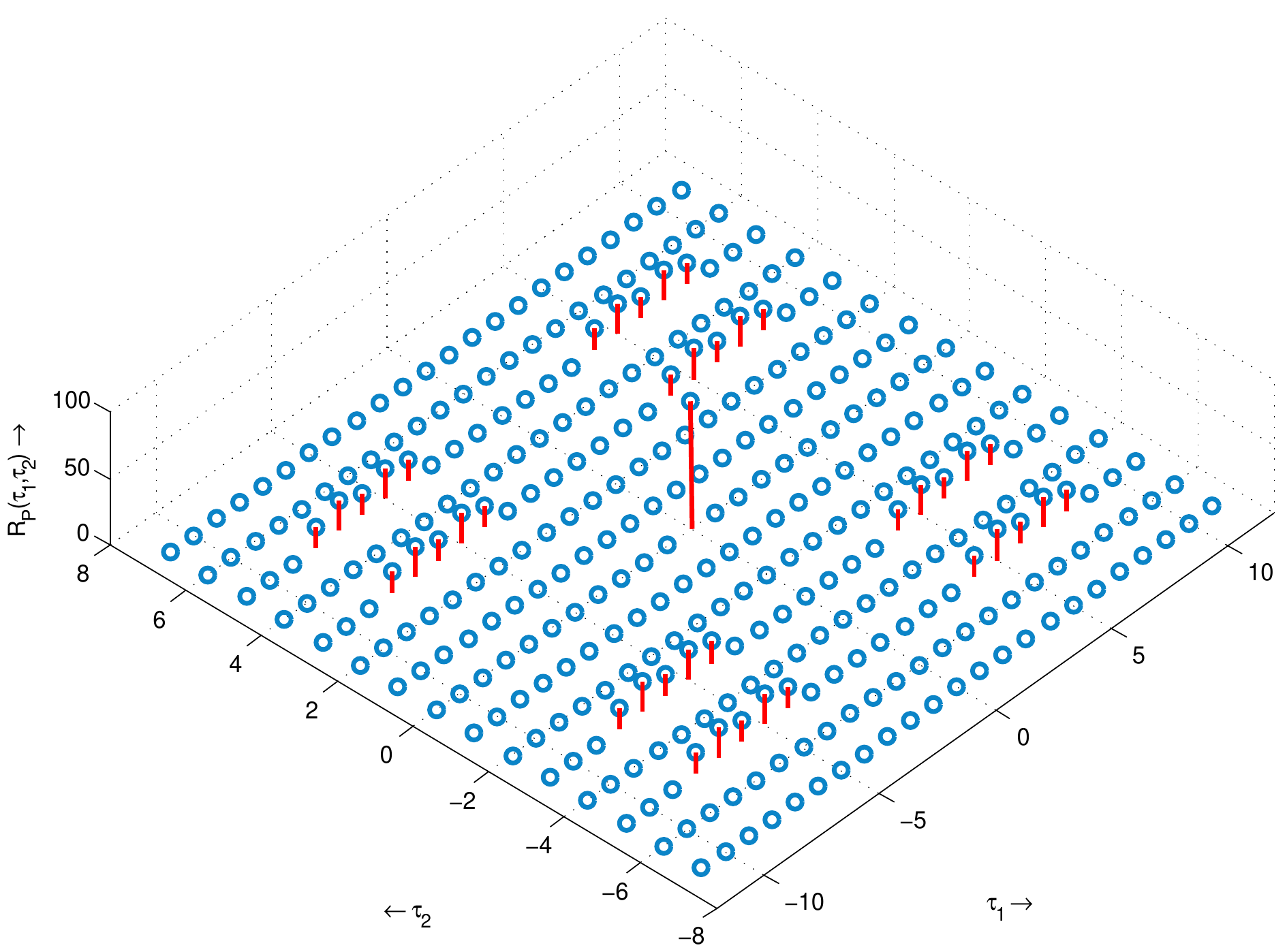}
		\caption{2D- periodic autocorrelation of $\mathcal{P}$}
		\label{fig:sub-first}
	\end{subfigure}
	\begin{subfigure}{.45\textwidth}
		\centering
		\includegraphics[width=\textwidth]{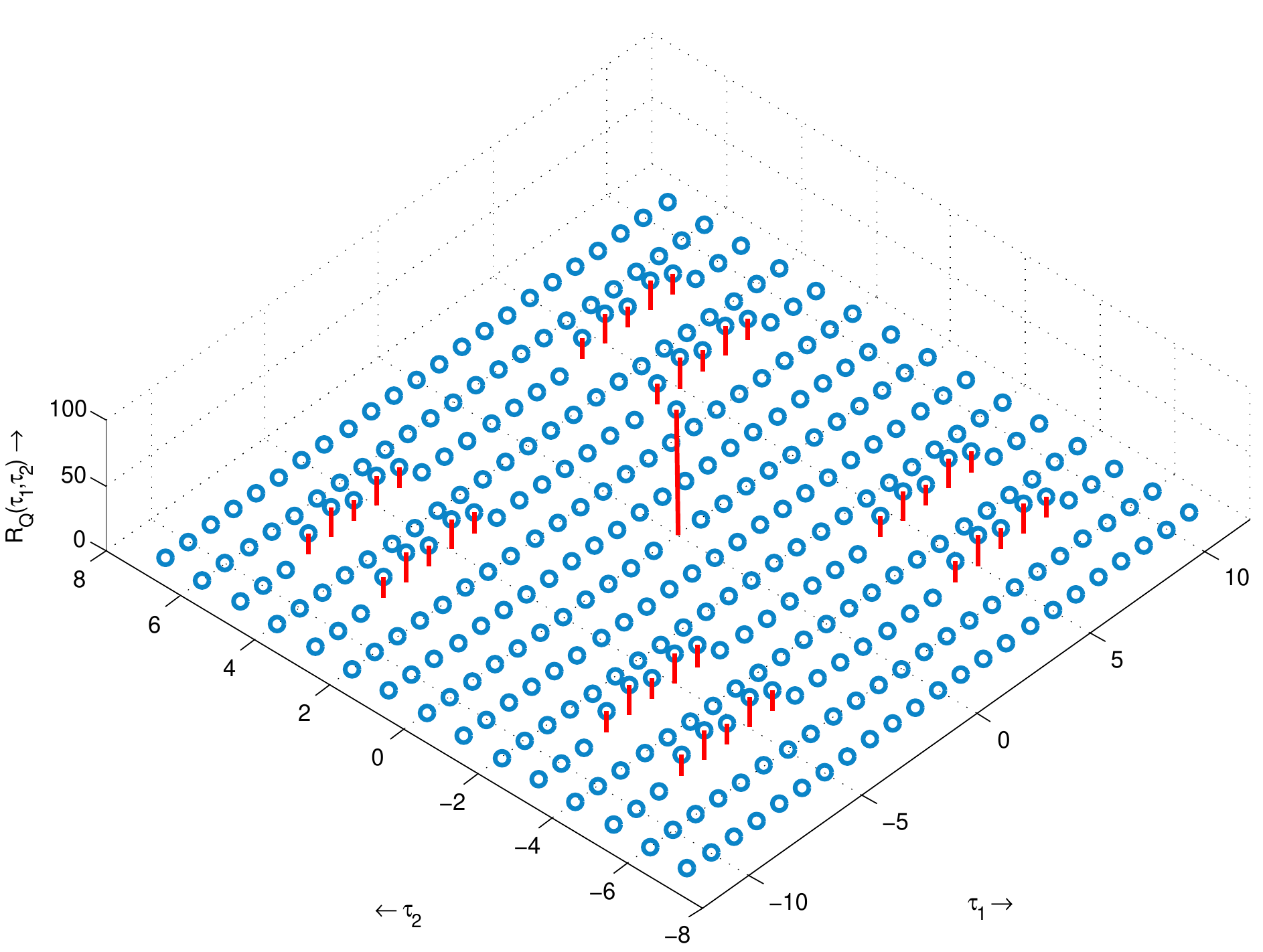}
		\caption{2D- periodic autocorrelation of $\mathcal{Q}$}
		\label{fig:sub-second}
	\end{subfigure}
	\begin{subfigure}{.45\textwidth}
		\centering
		\includegraphics[width=\textwidth]{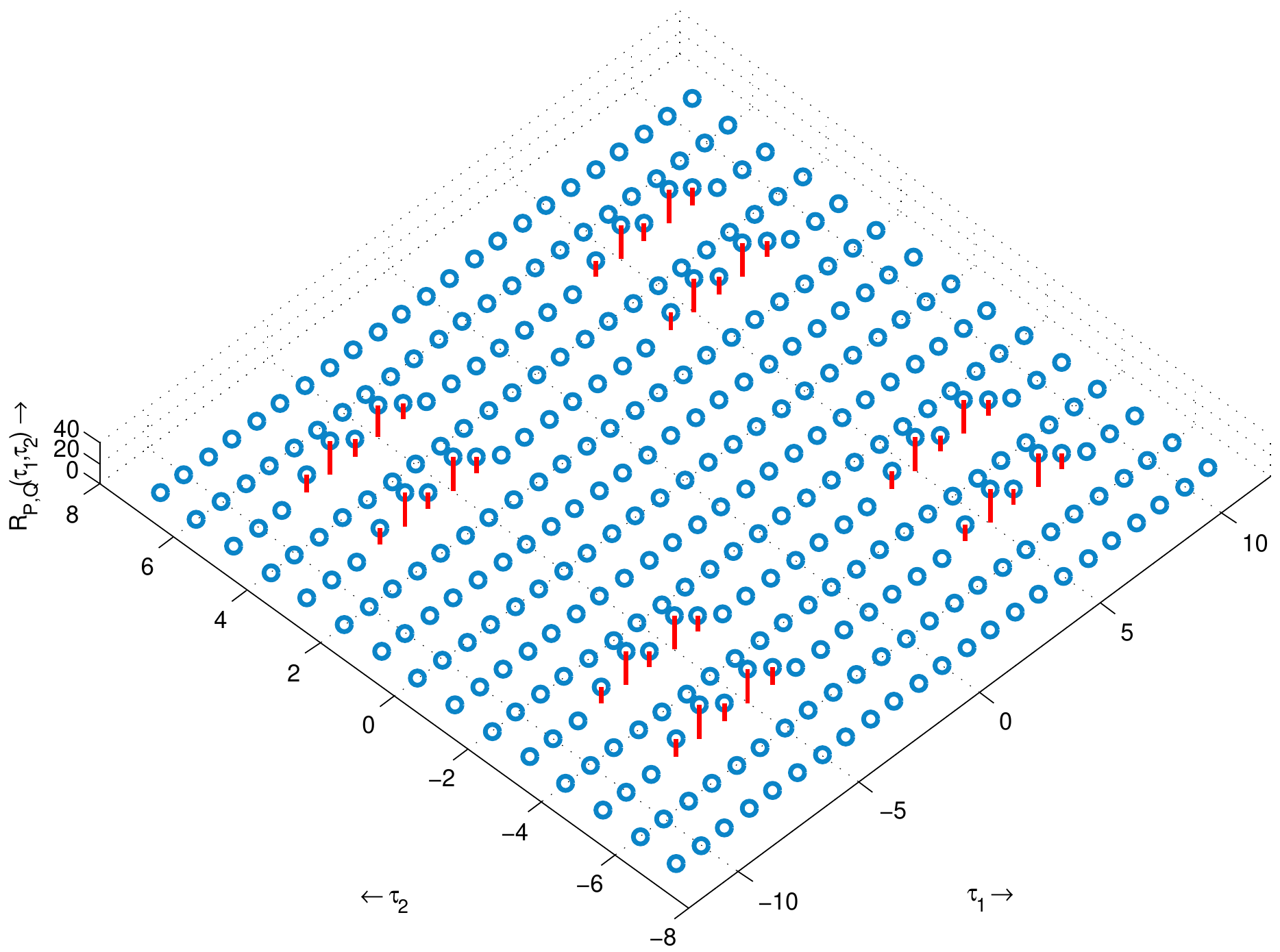}
		\caption{2D- periodic cross-correlation of $\mathcal{P}$ and $\mathcal{Q}$}
		\label{fig:sub-second}
	\end{subfigure}
	\caption{A glimpse of the 2D- periodic correlations of the GCAP given in Example \ref{ex3}.}
	\label{fig3}
\end{figure}
\end{example}
\section{Conclusion}\label{section 5}

In this paper, we have made two  contributions. Firstly, we have systematically constructed GCPs of lengths non-power-of-two where each of the constituent sequences have a periodic ZACZ and the pair have a periodic ZCCZ. This partially fills the gap left by the previous remarkable works of Gong \textit{et al.} and Chen \textit{et. al}, in terms of the lengths of the GCP. Secondly, by extending our construction to higher dimensions, we construct 2D- GCAPs. For the first time, the periodic 2D-autocorrelations of the constituent arrays and the periodic 2D-cross-correlation of the 2D-GCAP has been analysed. All the proposed 2D- GCAPs have large periodic 2D autocorrelation and cross-correlation zones. An interesting future
work will be to design Golay complementary sequence sets, consisting sequences of length non-power of two, which will have periodic ZACZ and ZCCZ around the in-phase position.


\begin{thebibliography}{10}
\providecommand{\url}[1]{#1}
\csname url@samestyle\endcsname
\providecommand{\newblock}{\relax}
\providecommand{\bibinfo}[2]{#2}
\providecommand{\BIBentrySTDinterwordspacing}{\spaceskip=0pt\relax}
\providecommand{\BIBentryALTinterwordstretchfactor}{4}
\providecommand{\BIBentryALTinterwordspacing}{\spaceskip=\fontdimen2\font plus
\BIBentryALTinterwordstretchfactor\fontdimen3\font minus
  \fontdimen4\font\relax}
\providecommand{\BIBforeignlanguage}[2]{{%
\expandafter\ifx\csname l@#1\endcsname\relax
\typeout{** WARNING: IEEEtran.bst: No hyphenation pattern has been}%
\typeout{** loaded for the language `#1'. Using the pattern for}%
\typeout{** the default language instead.}%
\else
\language=\csname l@#1\endcsname
\fi
#2}}
\providecommand{\BIBdecl}{\relax}
\BIBdecl

\bibitem{Golay51}
M.~J.~E. Golay, ``Static multislit spectrometry and its application to the
  panoramic display of infrared spectra,'' \emph{J. Opt. Soc. Am.}, vol.~41,
  no.~7, pp. 468--472, Jul. 1951.

\bibitem{Golay61}
M.~{Golay}, ``Complementary series,'' \emph{IRE Trans. Inf. Theory}, vol.~7,
  no.~2, pp. 82--87, Apr. 1961.

\bibitem{Davis1999}
J.~A. {Davis} and J.~{Jedwab}, ``Peak-to-mean power control in {OFDM}, {G}olay
  complementary sequences, and {R}eed-{M}uller codes,'' \emph{IEEE Trans. Inf.
  Theory}, vol.~45, no.~7, pp. 2397--2417, Nov. 1999.

\bibitem{Paterson2000}
K.~G. {Paterson}, ``Generalized {R}eed-{M}uller codes and power control in
  {OFDM} modulation,'' \emph{IEEE Trans. Inf. Theory}, vol.~46, no.~1, pp.
  104--120, Jan. 2000.

\bibitem{Georghiades2001}
C.~N. Georghiades, ``Complementary sequences for {ISI} channel estimation,''
  \emph{IEEE Trans. Inf. Theory}, pp. 1145--1152, 2001.

\bibitem{farrel2003}
K.~K. {Wong} and T.~{O'Farrell}, ``Application of complementary sequences in
  indoor wireless infrared communications,'' \emph{IEE Proc. -
  Optoelectronics}, vol. 150, no.~5, pp. 453--464, 2003.

\bibitem{abdi2007}
S.~{Wang} and A.~{Abdi}, ``{MIMO} {ISI} channel estimation using uncorrelated
  {G}olay complementary sets of polyphase sequences,'' \emph{IEEE Trans. Veh.
  Tech.}, vol.~56, no.~5, pp. 3024--3039, 2007.

\bibitem{lei2014}
C.~{Lei}, L.~{Zhigang}, C.~{Xiantao}, and L.~{Shaoqian}, ``Golay sequence based
  time-domain compensation of frequency-dependent {I/Q} imbalance,''
  \emph{China Commun.}, vol.~11, no.~6, pp. 1--11, 2014.

\bibitem{Spano1996}
E.~{Spano} and O.~{Ghebrebrhan}, ``Complementary sequences with high sidelobe
  suppression factors for {ST/MST} radar applications,'' \emph{IEEE Trans.
  Geosci. Remote Sens.}, vol.~34, no.~2, pp. 317--329, 1996.

\bibitem{Pezeshki2008}
A.~{Pezeshki}, A.~R. {Calderbank}, W.~{Moran}, and S.~D. {Howard}, ``Doppler
  resilient {G}olay complementary waveforms,'' \emph{IEEE Trans. Inf. Theory},
  vol.~54, no.~9, pp. 4254--4266, Sep. 2008.

\bibitem{Borwein2000}
P.~B. {Borwein} and R.~A. {Ferguson}, ``A complete description of {G}olay pairs
  for lengths up to 100,'' \emph{Mathematics of Computation}, vol.~73, p. 2003,
  2000.

\bibitem{Fan2007}
P.~{Fan}, W.~{Yuan}, and Y.~{Tu}, ``{Z}-complementary binary sequences,''
  \emph{IEEE Signal Process. Lett.}, vol.~14, no.~8, pp. 509--512, 2007.

\bibitem{Li2011}
X.~{Li}, P.~{Fan}, X.~{Tang}, and Y.~{Tu}, ``Existence of binary
  {Z}-complementary pairs,'' \emph{IEEE Signal Process. Lett.}, vol.~18, no.~1,
  pp. 63--66, 2011.

\bibitem{liu20141}
Z.~{Liu}, U.~{Parampalli}, and Y.~L. {Guan}, ``Optimal odd-length binary
  {Z}-complementary pairs,'' \emph{IEEE Trans. Inf. Theory}, vol.~60, no.~9,
  pp. 5768--5781, 2014.

\bibitem{liu2014}
Z.~{Liu}, U.~{Parampalli}, and Y.~L. {Guan}, ``On even-period binary {Z}-complementary pairs with large {ZCZ}s,''
  \emph{IEEE Signal Process. Lett.}, vol.~21, no.~3, pp. 284--287, 2014.

\bibitem{Adhikary2016}
A.~R. {Adhikary}, Z.~{Liu}, Y.~L. {Guan}, S.~{Majhi}, and S.~Z. {Budishin},
  ``Optimal binary periodic almost-complementary pairs,'' \emph{IEEE Signal
  Process. Lett.}, vol.~23, no.~12, pp. 1816--1820, 2016.

\bibitem{Adhikary2018}
A.~R. {Adhikary}, S.~{Majhi}, Z.~{Liu}, and Y.~L. {Guan}, ``New sets of
  even-length binary {Z}-complementary pairs with asymptotic {ZCZ} ratio of
  $3/4$,'' \emph{IEEE Signal Process. Lett.}, vol.~25, no.~7, pp. 970--973,
  2018.

\bibitem{Adhikary2020}
A.~R. {Adhikary}, S.~{Majhi}, Z.~{Liu}, and Y.~L. {Guan}, ``New sets of optimal odd-length binary {Z}-complementary pairs,''
  \emph{IEEE Trans. Inf. Theory}, vol.~66, no.~1, pp. 669--678, 2020.

\bibitem{chen2017}
C.~{Chen}, ``A novel construction of {Z}-complementary pairs based on
  generalized {B}oolean functions,'' \emph{IEEE Signal Process. Lett.},
  vol.~24, no.~7, pp. 987--990, 2017.

\bibitem{Adhikary20201}
A.~R. {Adhikary}, P.~{Sarkar}, and S.~{Majhi}, ``A direct construction of
  $q$-ary even length {Z}-complementary pairs using generalized {B}oolean
  functions,'' \emph{IEEE Signal Process. Lett.}, vol.~27, pp. 146--150, 2020.

\bibitem{Adhikary20191}
A.~R. {Adhikary} and S.~{Majhi}, ``New construction of optimal aperiodic
  {Z}-complementary sequence sets of odd-lengths,'' \emph{Electron. Lett.},
  vol.~55, no.~19, pp. 1043--1045, 2019.

\bibitem{chen20192}
C.~{Chen} and C.~{Pai}, ``Binary {Z}-complementary pairs with bounded
  peak-to-mean envelope power ratios,'' \emph{IEEE Commun. Lett.s}, vol.~23,
  no.~11, pp. 1899--1903, 2019.

\bibitem{Gong2013}
G.~Gong, F.~Huo, and Y.~Yang, ``Large zero autocorrelation zones of {G}olay
  sequences and their applications,'' \emph{IEEE Trans. commun.}, vol.~61,
  no.~9, pp. 3967--3979, 2013.

\bibitem{Chen201811}
C.~{Chen} and S.~{Wu}, ``Golay complementary sequence sets with large zero
  correlation zones,'' \emph{IEEE Trans. Commun.}, vol.~66, no.~11, pp.
  5197--5204, 2018.

\bibitem{Ohyama1978}
N.~Ohyama, T.~Honda, and J.~Tsujiuchi, ``An advanced coded imaging without side
  lobes,'' \emph{Opt. Commun.}, vol.~27, no.~3, pp. 339--344, 1978.

\bibitem{Luke1985}
H.~D. {L\"{u}ke}, ``Sets of one and higher dimensional {W}elti codes and
  complementary codes,'' \emph{IEEE Trans. Aerospace Electron. Syst.}, vol.
  AES-21, no.~2, pp. 170--179, March 1985.

\bibitem{bomer1990}
L.~{Bomer} and M.~{Antweiler}, ``Two-dimensional perfect binary arrays with 64
  elements,'' \emph{IEEE Trans. Inf. Theory}, vol.~36, no.~2, pp. 411--414,
  1990.

\bibitem{Jedwab20071}
J.~Jedwab and M.~G. Parker, ``Golay complementary array pairs,'' \emph{Des.
  Codes, Cryptogr.}, p. 2007.

\bibitem{Tang2001}
X.~H. {Tang}, P.~Z. {Fan}, D.~B. {Li}, and N.~{Suehiro}, ``Binary array set
  with zero correlation zone,'' \emph{Electron. Lett.}, vol.~37, no.~13, pp.
  841--842, 2001.

\bibitem{Hayashi2004}
T.~{Hayashi} and S.~{Okawa}, ``Binary array set having a cross-shaped
  zero-correlation zone,'' \emph{IEEE Signal Processing Letters}, vol.~11,
  no.~4, pp. 423--426, 2004.

\bibitem{Cheng2010}
C.~{Cheng}, T.~{Jiang}, and Y.~{Liu}, ``A novel class of {2-D} binary sequences
  with zero correlation zone,'' \emph{IEEE Signal Process. Lett.}, vol.~17,
  no.~3, pp. 301--304, 2010.

\bibitem{pai2019}
C.~{Pai}, Y.~{Ni}, Y.~{Liu}, M.~{Kuo}, and C.~{Chen}, ``Constructions of
  two-dimensional binary {Z}-complementary array pairs,'' in \emph{IEEE Int.
  Symp. Inf. Theory (ISIT)}, 2019, pp. 2264--2268.

\bibitem{Matsufuji2010}
S.~Matsufuji and T.~Matsumoto, ``On logic functions of complementary arrays of
  length $2^n$,'' in \emph{Proc. 10th WSEAS int. conf. appl. info. commun., and
  3rd WSEAS int. conf. Biomed. electron. .biomed. info.}\hskip 1em plus 0.5em
  minus 0.4em\relax World Scientific and Engineering Academy and Society
  (WSEAS), 2010, pp. 208--215.

\end{thebibliography}

\end{document}